\documentclass[twocolumn]{aastex62}

\usepackage{graphicx,amsmath}
\usepackage{amssymb}
\usepackage{subfigure}
\usepackage{color}
\usepackage{bm}
\usepackage{soul}
\usepackage{CJK}
\hypersetup{linkcolor=blue, citecolor=blue}

\newcommand{\hi}{H\thinspace{\sc i}}

\newcommand{\htwo}{H$_2$}

\newcommand{\mhi}{$M_{\rm HI}$}

\newcommand{\Msol}{\hbox{\thinspace $M_{\odot}$}}
\newcommand{\kms}{\,km\,s$^{-1}$}

\received{2019 August 1}
\revised{2019 September 9}
\accepted{2019 October 3}

\shorttitle{Massive quiescent disk galaxies are HI-rich}
\shortauthors{Zhang et al.}

\begin{document}

%\title{NEARLY ALL MASSIVE QUIESCENT DISK GALAXIES HAVE A SURPRISINGLY LARGE ATOMIC GAS RESERVOIR}
\title{{\bf \Large Nearly all Massive Quiescent Disk Galaxies have a Surprisingly Large \\Atomic Gas Reservoir}}

\correspondingauthor{Yingjie Peng}
\email{yjpeng@pku.edu.cn }

\author{Chengpeng Zhang}
\affil{Kavli Institute for Astronomy and Astrophysics, Peking University, 5 Yiheyuan Road, Beijing 100871, China}
\affil{Department of Astronomy, School of Physics, Peking University, 5 Yiheyuan Road, Beijing 100871, China}

\author{Yingjie Peng}
\affiliation{Kavli Institute for Astronomy and Astrophysics, Peking University, 5 Yiheyuan Road, Beijing 100871, China}

\author{Luis C. Ho}
\affil{Kavli Institute for Astronomy and Astrophysics, Peking University, 5 Yiheyuan Road, Beijing 100871, China}
\affil{Department of Astronomy, School of Physics, Peking University, 5 Yiheyuan Road, Beijing 100871, China}

\author{Roberto Maiolino}
\affiliation{Cavendish Laboratory, University of Cambridge, 19 J. J. Thomson Avenue, Cambridge CB3 0HE, UK}
\affiliation{Kavli Institute for Cosmology, University of Cambridge, Madingley Road, Cambridge CB3 0HA, UK}

\author{Avishai Dekel}
\affiliation{Racah Institute of Physics, The Hebrew University, Jerusalem 91904, Israel}
\affiliation{SCIPP, University of California, Santa Cruz, CA 95064, USA}

\author{Qi Guo}
\affiliation{Key Laboratory for Computational Astrophysics, National Astronomical Observatories, \\Chinese Academy of Sciences, Beijing 100012, China}
\affiliation{School of Astronomy and Space Science, University of Chinese Academy of Sciences, Beijing 100049, China}

\author{Filippo Mannucci}
\affiliation{Istituto Nazionale di Astrofisica, Osservatorio Astrofisico di Arcetri, Largo Enrico Fermi 5, I-50125 Firenze, Italy}

\author{Di Li}
\affiliation{CAS Key Laboratory of FAST, National Astronomical Observatories, Chinese Academy of Sciences, Beijing 100012, China}
\affiliation{School of Astronomy and Space Science, University of Chinese Academy of Sciences, Beijing 100049, China}

\author{Feng Yuan}
\affiliation{Key Laboratory for Research in Galaxies and Cosmology, Shanghai Astronomical Observatory, Chinese Academy of Sciences, 80 Nandan Road, Shanghai 200030, China}

\author{Alvio Renzini}
\affiliation{INAF - Osservatorio Astronomico di Padova, Vicolo dell'Osservatorio 5, I-35122 Padova, Italy}

\author{Jing Dou}
\affil{Kavli Institute for Astronomy and Astrophysics, Peking University, 5 Yiheyuan Road, Beijing 100871, China}
\affil{Department of Astronomy, School of Physics, Peking University, 5 Yiheyuan Road, Beijing 100871, China}

\author{Kexin Guo}
\affil{Kavli Institute for Astronomy and Astrophysics, Peking University, 5 Yiheyuan Road, Beijing 100871, China}
\affil{International Centre for Radio Astronomy Research, University of Western Australia, Crawley, WA 6009, Australia}

\author{Zhongyi Man}
\affil{Kavli Institute for Astronomy and Astrophysics, Peking University, 5 Yiheyuan Road, Beijing 100871, China}
\affil{Department of Astronomy, School of Physics, Peking University, 5 Yiheyuan Road, Beijing 100871, China}

\author{Qiong Li}
\affil{Kavli Institute for Astronomy and Astrophysics, Peking University, 5 Yiheyuan Road, Beijing 100871, China}
\affil{Department of Astronomy, School of Physics, Peking University, 5 Yiheyuan Road, Beijing 100871, China}

\begin{abstract}   

\noindent The massive galaxy population above the characteristic Schechter mass $M_*\approx 10^{10.6}\Msol$ contributes to about half of the total stellar mass in the local Universe. These massive galaxies usually reside in hot dark matter haloes above the critical shock-heating mass $\sim$$10^{12}\Msol$, where the external cold gas supply to these galaxies is expected to be suppressed. When galaxies run out of their cold gas reservoir, they become dead and quiescent. Therefore, massive quiescent galaxies living in hot haloes are commonly believed to be gas-poor. Based on the data from SDSS, ALFALFA, GASS and COLD GASS surveys, here we show that the vast majority of the massive, quiescent, central disk galaxies in the nearby Universe have a remarkably large amount of cold atomic hydrogen gas, surprisingly similar to star-forming galaxies. Both star-forming and quiescent disk galaxies show identical symmetric double-horn \hi~spectra, indicating similar regularly rotating \hi~disks. Relative to their star-forming counterparts, massive quiescent central disk galaxies are quenched because of their significantly reduced molecular gas content, lower dust content, and lower star formation efficiency. Our findings reveal a new picture, which clearly demonstrates the detailed star-formation quenching process in massive galaxies and provides a stringent constraint on the physical mechanism of quenching.

\end{abstract}


\section{Introduction} \label{sec:intro}

Identifying the physical mechanism responsible for star formation quenching is one of the most debated open questions. In general, the level of star formation in the galaxy is controlled by its cold gas reservoir and star formation efficiency. Investigating the cold gas content in galaxies will provide direct observational evidence of how quenching may happen.

On average, the cold gas reservoir of quiescent galaxies was found to be significantly less than that of star-forming galaxies \citep[e.g.,][]{Fabello2011self,Huang2012,Brown2015,Saintonge2016self,Tacconi2018self,Catinella2018self}. However, some quiescent galaxies were found to have a large amount of \hi~gas content \citep[e.g.,][]{Schommer1983Aself,Gereb2016self,Gereb2018self,George2019self,Parkash2019self}. On the other hand, although the majority of quiescent galaxies are ellipticals or lenticulars \citep{Emsellem2011self}, there exists a significant population of red spiral galaxies \citep[e.g.,][]{Masters2010,Tojeiro2013}. Various physical mechanisms for quenching star formation are currently entertained, such as AGN feedback \citep{Croton2006,Fabian2012,Harrison2017a}, environmental effects \citep{Kauffmann2004,Baldry2006,Peng2012}, major merger \citep{Mihos1996self,Hopkins2008self}, halo quenching \citep{Dekel2006self}, morphological quenching \citep{Martig2009}, gravitational quenching \citep{Genzel2014} and strangulation \citep{Larson1980,Peng2015}. The morphological differences among quenched galaxies may be related to the specific mechanism that was responsible for their quenching.

Investigating the cold gas content in different quiescent galaxies, such as, central/satellite, disk/elliptical galaxies, may provide stringent constraints on different quenching mechanisms. In this study, we focus on the cold gas content in central galaxies above the characteristic Schechter mass $M_*\approx 10^{10.6}\Msol$. These massive galaxies usually reside in hot dark matter haloes above the critical shock-heating mass $M_{\rm shock}\approx 10^{12}\Msol$ \citep{Dekel2006self}, where the external cold gas supply to these galaxies is expected to be suppressed \citep{Dekel2006self,Dekel2008self}. Using the data from multi-wavelength sky surveys, here we show that the vast majority of the massive quiescent central galaxies with disk morphologies still have a large amount of cold atomic hydrogen gas, surprisingly similar to star-forming galaxies. We further show that the reason of the low-level star formation in these galaxies is the significant lower molecular gas mass and lower star formation efficiency of the molecular gas. The \citet{Chabrier2003self} initial mass function (IMF) is used throughout this work and we assume the following cosmological parameters: $\Omega_m=0.3, \Omega_\Lambda=0.7, H_0=70\,\rm {km\,s^{-1} Mpc^{-1}}$.

\section{Sample} \label{sec:sample} 
\subsection{SDSS}
The parent galaxy sample analyzed in this paper is the same Sloan Digital Sky Survey \cite[SDSS,][]{Abazajian2009} DR7 sample that we constructed in \citet{Peng2010,Peng2012,Peng2015}. The parent photometric sample contains 1,579,314 objects after removing duplicates, of which 72,697 have reliable spectroscopic redshift measurements in the redshift range $0.02<z<0.05$. 

The stellar masses ($M_*$) for the SDSS galaxies are determined from the $k$-correction program v4\_1\_4 \citep{Blanton2007} with stellar population synthesis models of \citet{Bruzual2003}. The derived stellar masses are highly consistent with the published stellar masses of \citet{Kauffmann2003} with a small difference of $\sim$0.1\,dex. The star formation rates (SFRs) are taken from the value-added MPA-JHU SDSS DR7 catalogue \citep{Brinchmann2004} and converted to Chabrier IMF. These SFRs are based on H$\alpha$ emission-line luminosities, corrected for extinction using the H$\alpha$/H$\beta$ ratio. To correct for the aperture effects, the SFRs outside the SDSS 3\arcsec~fiber were obtained by performing the spectral energy distribution (SED) fitting to the $ugriz$ photometry outside the fiber, using the models and methods described in \citet{Salim2007}. Since the H$\alpha$ emission of AGN and composite galaxies are likely to be contaminated by their nuclear activity, their SFRs are derived based on the strength of the 4000\,$\mbox{\normalfont\AA}$ break as calibrated with H$\alpha$ for non-AGN, pure star-forming galaxies \citep[see details in][]{Brinchmann2004}.

We classify our sample into central galaxies and satellite galaxies using the SDSS DR7 group catalogue from \citet{Yang2005self} and \citet{Yang2007}. To reduce the contamination of the central sample by spurious interlopers into the group, we define central galaxies to be simultaneously both the most massive and the most luminous (in $r$-band) galaxy within a given group. The centrals also include single galaxies that do not have identified companions above the SDSS flux limit.

The morphology classifications are from the Galaxy Zoo (GZ) project \citep{Lintott2011a}. In GZ, the image of each galaxy was viewed and classified by dozens of volunteers and a morphology flag (``spiral'', ``elliptical'' or ``uncertain'') is assigned to each galaxy after a de-biasing process. Most lenticular/S0 galaxies with smooth and rounded profiles are classified in GZ as ``elliptical'' or ``uncertain''.  Since ``spirals'' in GZ include disk galaxies with or without clear spiral arms, we simply designate all galaxies classified as ``spiral'' in GZ as ``disk'' galaxies. In total, 3\% of these galaxies are excluded because their P\_MG values in GZ are greater than 0.3, which indicates that they are very likely mergers.

\subsection{ALFALFA and sample matching} \label{AA}

The main \hi~sample used in this paper is from the Arecibo Legacy Fast ALFA (ALFALFA) survey \citep{Haynes2011,Haynes2018self}. We use the 100\% ALFALFA extragalactic \hi~source catalogue, which contains 31,502 \hi~detections. Both of ``code I'' and ``code II'' detections \citep[see details in][]{Haynes2011} are used in our work. The final \hi~sample contains 14,640 sources in redshift range of $0.02-0.05$ in the ALFALFA-SDSS overlap region ($\sim$4000 deg$^2$).

The most probable optical counterpart (OC) of each \hi~detection has been assigned in the ALFALFA catalogue. The ALFALFA \hi~detections and SDSS galaxies are then cross-matched using the following criteria: (1) the spatial separation between the OC and SDSS galaxy is less than 5 arcsec; (2) the velocity difference between the \hi~source and SDSS galaxy is less than 300\kms. With these selection criteria, the matched sample consists of 10,972 galaxies in the redshift range of $0.02 - 0.05$. 

The measured \hi~spectra may be contaminated by close companions due to the large Arecibo beam ($\sim$4\,arcmin).  Multiple SDSS galaxies within the beam radius and within a velocity difference of 3 times the \hi~line width ($W_{\rm 50}$) were excluded as potential \hi~contaminants. About 1,300 galaxies were removed.  Including these galaxies with ambiguous \hi~measurements produces negligible changes to the results presented in this paper. The final SDSS-ALFALFA matched sample used in our analysis contains 9,595 galaxies.

Both of the SDSS spectroscopy sample and ALFALFA sample are from flux-limited surveys. Faint sources are progressively missed at higher redshifts. The SDSS spectroscopic selection $r<17.77$ is roughly complete at $z = 0.05$ above a stellar mass of $\sim$$10^{9.2}$\Msol~for star-forming galaxies; for passive galaxies, the corresponding limit is $\sim$$10^{9.8}$\Msol. The ALFALFA sample is approximately complete at $z = 0.05$ above an \hi~gas mass of $\sim$$10^{9.8}$\Msol~for $W_{\rm 50}$ = 100\kms~and of $\sim$$10^{10.2}$\Msol~for $W_{\rm 50}$ = 400\kms. To include the population of galaxies with lower stellar masses and \hi~masses in our analysis,  we corrected our sample by using the ``$V_{\rm max}$ method''. In detail,  we calculated the maximum redshift at which the galaxy can still be detected according to the ALFALFA and SDSS sensitivity limits. The maximum redshift was then used to calculate the observable maximum co-moving volume ($V_{\rm max}$) for each galaxy. By assuming the spatial distribution of our sample is homogenous in the co-moving space, we weight each galaxy using the value of $V_{\rm total}/V_{\rm max}$ to account for the galaxies missed in the surveys, where $V_{\rm total}$ is the total co-moving volume that our sample spans. The corrections of SDSS and ALFALFA sample are performed independently and they are combined together to correct the matched sample.

\begin{figure}[htbp]
    \begin{center}
       \includegraphics[width=\columnwidth]{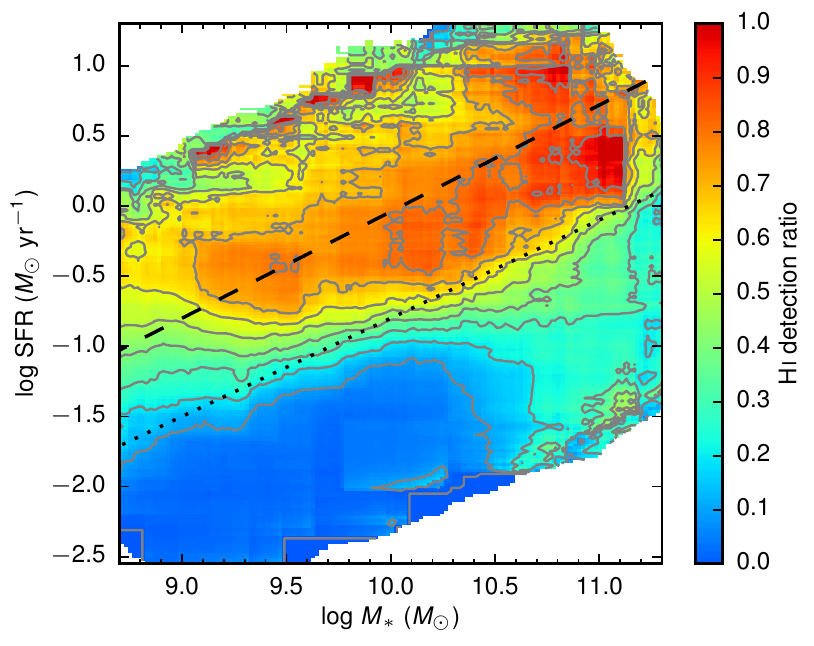}
    \end{center}
\caption{\hi~detection ratio in ALFALFA as a function of stellar mass ($M_*$) and star formation rate (SFR), determined within moving boxes of size 0.5\,dex in mass and 0.5\,dex in SFR. Each galaxy is weighted by a correction factor to account for selection effects. The dashed line indicates the position of the star-forming main sequence defined in \citet{Renzini2015}. The dotted line indicates the approximate divide between star-forming and quiescent galaxies according to their bimodal distribution in the SFR-$M_*$ plane.}
 \label{hi_2d}
\end{figure}
%---------------------------------------------------------------------------------------------------------------------

\begin{figure*}[htbp]
    \begin{center}
       \includegraphics[width=182mm]{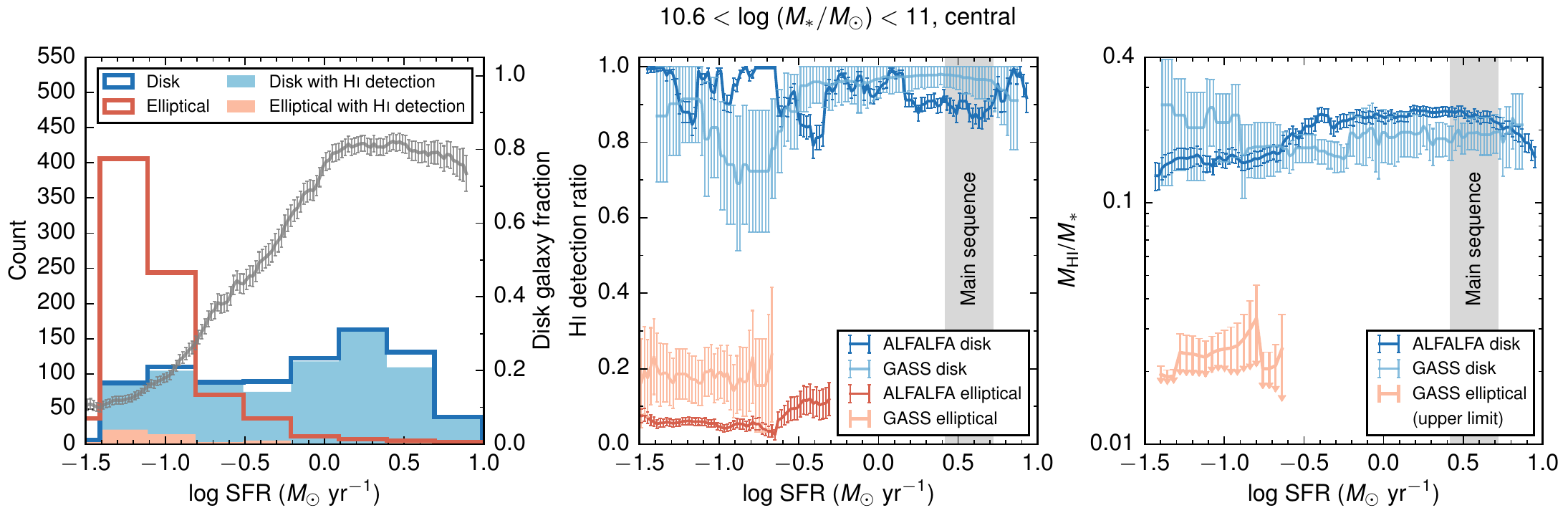}
    \end{center}
\caption{\textbf{Left,} SFR distribution function for central disk galaxies (blue line) and central elliptical galaxies (red line) in the narrow stellar mass range of $10^{10.6} - 10^{11}\Msol$. The shaded regions show the SFR distribution of galaxies with \hi~detection, in blue for central disks and in red for central ellipticals. About 30\% of the galaxies with uncertain morphology are not shown here. The grey line shows the fractional abundance of disk galaxies, obtained by using a sliding average of 0.5\,dex in SFR. Error bars are derived from the binomial error of the fraction with 68\% confidence level. \textbf{Middle,} \hi~detection ratio in ALFALFA and GASS as a function of SFR for central disk galaxies (blue lines) and central elliptical galaxies (red lines) in the stellar mass range of $10^{10.6} - 10^{11}\Msol$. These values are determined by using the sliding average of 0.5\,dex in SFR, and error bars are derived from the binomial error of the fraction with 68\% confidence level. \textbf{Right,} \hi~gas mass to stellar mass ratio (\mhi/$M_*$) for central disk galaxies and central elliptical galaxies (upper limit) in the same stellar mass range. For elliptical galaxies in GASS, the upper limits of non-detections are included in the average so that it is plotted as pink arrows. The error bars on the lines indicate the 1\,$\sigma$ uncertainty around the mean value. The grey shades indicate the position of the star-forming main sequence defined in \citet{Renzini2015}.}
 \label{hi}
\end{figure*}
%---------------------------------------------------------------------------------------------------------------------

%---------------------------------------------------------------------------------------------------------------------

\vspace{-0.2cm}
\subsection{GASS and COLD GASS}
The data from a deeper \hi~survey, $GALEX$  Arecibo SDSS  Survey \citep[GASS,][]{Catinella2013}, is also used in this work. The representative GASS sample includes 47 central disk galaxies and 43 central elliptical galaxies in the stellar mass range of $10^{10.6}-10^{11}\Msol$. A randomly selected subset of GASS parent sample were observed in CO(1-0) in the COLD GASS survey using the IRAM 30m telescope \citep{Saintonge2011}, which is used for the study of molecular gas in this work. The COLD GASS sample includes 28 central disk galaxies in the stellar mass range of $10^{10.6} - 10^{11}\Msol$.

For the massive galaxies concerned in this paper, a detection limit of $M_{\rm HI}/M_*\approx 0.015$ was reached in GASS survey and $M_{\rm H_2}/M_*\approx 0.015$ was reached in COLD GASS survey. To correct the flat log\,$M_*$ distribution of these samples, each galaxy was weighted according to its stellar mass such that the weighted log\,$M_*$ distribution is similar to that of all SDSS galaxies.

%---------------------------------------------------------------------------------------------------------------------
\begin{figure*}[t]
    \begin{center}
       \includegraphics[width=178mm]{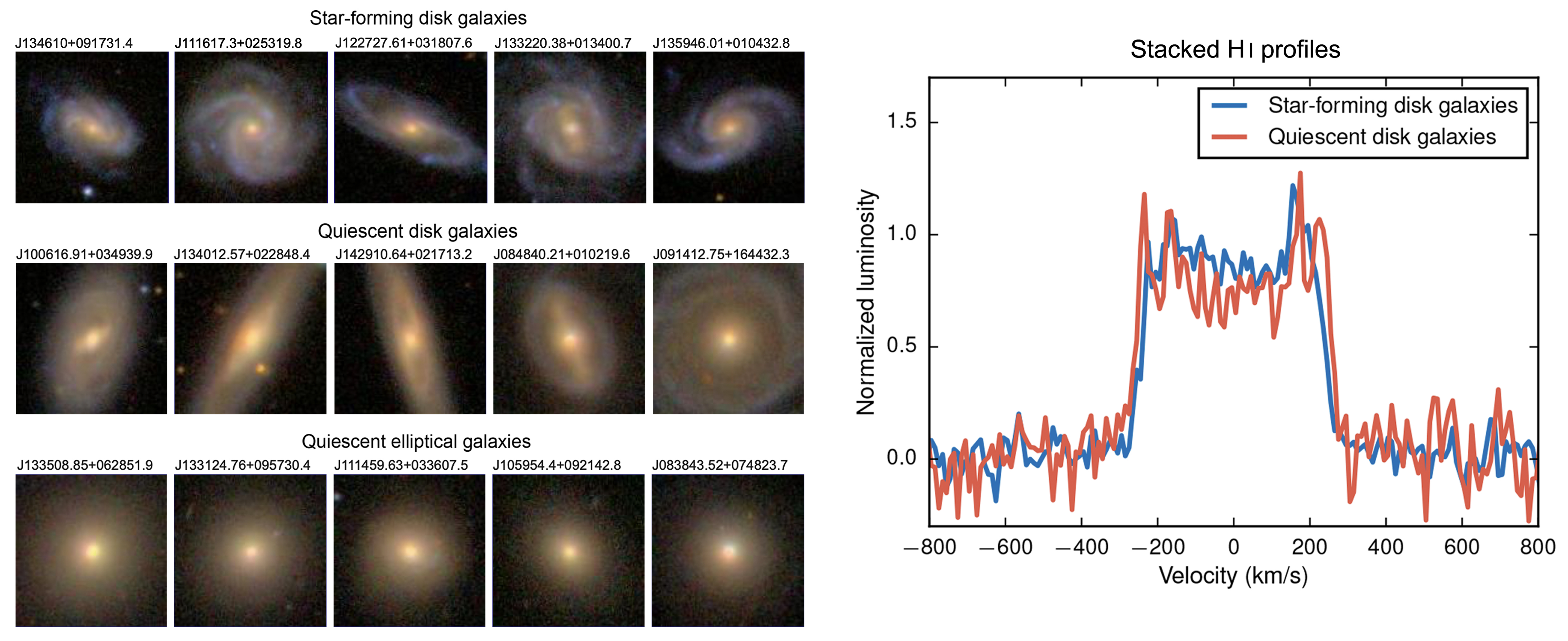}
    \end{center}
\caption{\textbf{Left,} SDSS $gri$ composite images of randomly selected central galaxies in the stellar mass range of $10^{10.6}-10^{11}\Msol$ for star-forming disk galaxies (top), quiescent disk galaxies (middle) and quiescent elliptical galaxies (bottom). The size of each image is 50 $\times$ 50 arcsec$^2$. \textbf{Right,} Stacked \hi~profiles for star-forming and quiescent central disk galaxies in ALFALFA. Galaxies with inclination angle $i>60^{\circ}$ (nearly edge-on) in the stellar mass range of $10^{10.6}-10^{11}\Msol$ are selected for stacking. Before stacking, each spectrum is normalized by the integrated luminosity, and the velocity of each channel is divided by a factor of sin $i$ to correct the projection effect. Star-forming and quiescent galaxies are divided by the dotted line in Figure \ref{hi_2d}.}
 \label{image}
\end{figure*}
%---------------------------------------------------------------------------------------------------------------------

%---------------------------------------------------------------------------------------------------------------------
\section{Results}

\subsection{Atomic hydrogen gas} \label{sec:hi}

With the $V_{\rm max}$ correction described in Section \ref{AA}, we calculated the \hi~detection ratio for SDSS galaxies in the SDSS-ALFALFA overlap region in the redshift range of $0.02-0.05$. The value of the corrected \hi~detection ratio in ALFALFA can be regarded as the fraction of galaxies with \mhi\,$>10^{9.3}$\,\Msol, which is the ALFALFA detection limit at $z\sim0.02$ for typical massive galaxies with \hi~line width of $300$\kms.

Figure \ref{hi_2d} shows the \hi~detection ratio in the SFR-$M_*$ plane for all SDSS galaxies. On average the \hi~detection ratio is above 70\% for star-forming galaxies on the main sequence, and it drops rapidly toward the passive sequence (below the dotted line).  The very low \hi~detection ratio (less than 10\% on average) for quiescent galaxies suggests that the quenching process may significantly impact the \hi~gas content in galaxies. However, it is surprising that \hi~is also detected in about 30\% of the massive quiescent galaxies with stellar mass above the Schechter characteristic mass $M_*\approx 10^{10.6}\Msol$. What makes them different from the other 70\% of massive quiescent galaxies with no \hi~detection? After a thorough examination of all observable parameters against these two populations of massive quiescent galaxies with and without \hi~detection, we find that almost all quiescent disk galaxies have \hi~detections, while quiescent ellipticals do not. 

Since the extended \hi~gas is very sensitive to environmental effects \citep[e.g.,][]{Giovanelli1985,Catinella2013}, we only focus on central galaxies in the following analyses. We also selected a narrow stellar mass bin of $10^{10.6}-10^{11}\Msol$ to minimize the effect of the dependence of SFR and gas mass on stellar mass. The left panel of Figure \ref{hi} shows the SFR distribution of central disk galaxies (blue line) and central elliptical galaxies (red  line) in this stellar mass range and the grey line shows the fraction of disks relative to all galaxies (including 30\% of the galaxies with uncertain morphology). It is well known that at a given stellar mass the distribution of SFR is bimodal \citep{Kauffmann2003, Baldry2006}. However, once we separate the galaxy population into disk galaxies and elliptical galaxies, the bimodality disappears. Although the quiescent galaxies are dominated by ellipticals, there is a substantial population of quenched disk galaxies.

The shaded regions in the left panel of Figure \ref{hi} show the SFR distribution of galaxies with \hi~detection. It is clear that almost all central disks have \hi~detection in ALFALFA from star-forming ones to the quiescents, while only few central ellipticals have \hi~detection. The  \hi~detection ratio of these galaxies are then quantified in the middle panel of Figure \ref{hi}. From the star-forming main sequence to the passive sequence, the average \hi~detection ratio for central disk galaxies is $\sim$90\% in ALFALFA and also in the deeper GASS survey, which implies that nearly all massive quiescent central disk galaxies have significant amounts of \hi~gas, e.g.,  with an \hi~gas fraction of $M_{\rm HI}/M_*>0.1$.

The right panel of Figure \ref{hi} shows the average \hi~gas mass to stellar mass ratio (\mhi/$M_*$) as a function of SFR for central disk galaxies. In ALFALFA, each galaxy has been weighted using the ``$V_{\rm max}$ method'' to correct the selection bias when deriving the average value. With this correction, our central disk galaxy sample has a very high \hi~detection ratio of $\sim$90\%, hence the \hi-detected central disk galaxies can be used as a fair representation of the whole sample of central disk galaxies. Encouragingly, the derived average value of \mhi/$M_*$ from ALFALFA (dark blue line in the right panel of Figure \ref{hi}) is very similar to the one derived from GASS (light blue line), which is a deeper representative sample (though with a much smaller sample size). Figure \ref{hi} clearly shows that, when SFR drops progressively from the star-forming main sequence to the passive sequence, the average \hi~gas mass of central disk galaxies remains almost the same. Massive quiescent central disk galaxies are surprising as abundant in \hi~gas as star-forming galaxies. 

As shown by the red lines in Figure \ref{hi}, the \hi~detection ratio for central ellipticals is only 10\%$-$20\%. In GASS survey, the upper limits of \mhi/$M_*$ for the non-detections are well constrained (down to 0.015). Thus, we calculate the average value of \mhi/$M_*$ for central ellipticals by including the upper limits of non-detections in GASS, as shown by the pink arrows. The average \hi~gas mass in central ellipticals is much lower than that of quiescent central disk galaxies. The very low \hi~gas amount in the overall quiescent population shown in previous literatures are due to the increasing fraction of elliptical galaxies when SFR decreases.

The SDSS $gri$ composite images of randomly selected examples of the star-forming disks, quiescent disks and quiescent ellipticals are shown in Figure \ref{image}. We inspected all of the \hi~spectra for central disk galaxies in the stellar mass range of $10^{10.6}-10^{11}\Msol$ in ALFALFA. Similar to star-forming ones, most quiescent central disk galaxies show characteristically symmetric double-horn \hi~profiles, indicating regularly rotating \hi~disks with little significant kinematic perturbations or contributions from companions, extra-planar gas, or tidal tails. As shown in the right panel of Figure \ref{image}, the stacked \hi~spectra for star-forming and quiescent disk galaxies show little difference.

%-------------------------------------------------------------------------------------------------------------------
\subsection{Molecular hydrogen gas and dust} \label{sec:h2}

%---------------------------------------------------------------------------------------------------------------------
\begin{figure*}[htbp]
    \begin{center}
       \includegraphics[width=180mm]{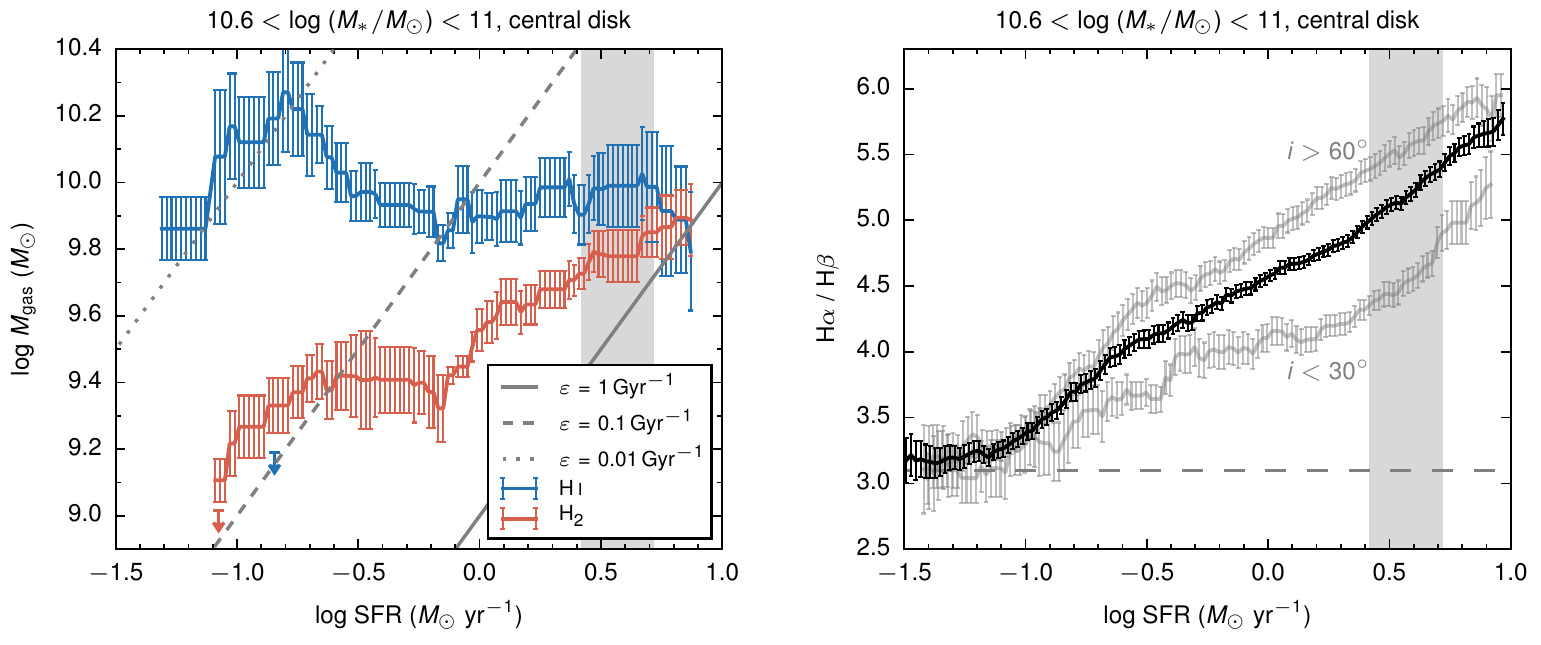}
    \end{center}
\caption{\textbf{Left,} The average atomic gas mass (in blue) and molecular gas mass (in red) as a function of SFR, from star-forming sequence to passive sequence, for central disk galaxies in the stellar mass range $10^{10.6} - 10^{11}\Msol$ in COLD GASS. The narrow stellar mass bin is used to minimize the effect of dependence of SFR and gas mass on stellar mass. The blue arrow shows the \hi~mass upper limit of the only galaxy without \hi~detection and the red arrow shows the \htwo~mass upper limit of the only galaxy without \htwo~detection. These non-detections are not included to derive the average gas mass. The grey lines indicate the constant star formation efficiency for three different values. \textbf{Right,} The average value of Balmer decrement (H$\alpha$/H$\beta$) for all central disk galaxies (black line), for those with $i<30^{\circ}$ (nearly face-on, lower grey line),  and for those with $i>60^{\circ}$ (nearly edge-on, upper grey line), within the same stellar mass range in SDSS. The horizontal dash line indicates the intrinsic value of 3.1 without dust extinction. In each panel, the grey shade indicates the position of the star-forming main sequence. The average values are calculated in the sliding box of 0.5\,dex in SFR; error bars on each line indicate the 1\,$\sigma$ uncertainty around the mean value. }
 \label{h2}
\end{figure*}
%---------------------------------------------------------------------------------------------------------------------

The substantial \hi~gas amount in massive quiescent central disk galaxies indicates that these quenched galaxies have enough raw material to fuel star formation.  The low level of SFR of these galaxies hence requires that the \hi~gas cannot be effectively converted into \htwo, and/or the \htwo~gas has very low star formation efficiency ($\varepsilon_{\rm H_2}$ = SFR/$M_{\rm H_2}$). Although it is well known that quiescent galaxies have lower molecular gas mass and $\varepsilon_{\rm H_2}$ on average compared to star-forming ones \citep[e.g.,][]{Saintonge2012self,Tacconi2018self}, it remains unclear what if we only select central disk galaxies.

The left panel of Figure \ref{h2} shows the mean \hi~(in blue) and \htwo~gas mass (in red) as a function of SFR for the central disk galaxies in the stellar mass range of $10^{10.6} - 10^{11}\Msol$ in COLD GASS sample. The upper limits of the only galaxy without \hi~detection and the only galaxy without \htwo~detection are shown as arrows and are not included to derive the average gas mass. As SFR decreases from the star-forming sequence to the passive sequence, the mean \hi~gas mass remains largely the same, while the mean \htwo~gas mass decreases by a factor of $\sim$10. The grey lines mark three different constant values of the star formation efficiency. As SFR drops, $\varepsilon_{\rm H_2}$ also drops by about a factor of 10, from 1\,Gyr$^{-1}$ to 0.1\,Gyr$^{-1}$.

The dust content provides an alternative estimate of the gas mass in galaxies \citep[e.g.,][]{Leroy2011self}. The Balmer decrement (H$\alpha$/H$\beta$) can be used as a proxy for dust content \citep{Kennicutt1992self} inside the SDSS fiber of 3\,arcsec, corresponding to $1.2-3$ kpc in the redshift range $z  = 0.02 - 0.05$. The black line in the right panel of Figure \ref{h2} shows that the mean value of the Balmer decrement for these galaxies drops rapidly with decreasing SFR. Since the dust attenuation depends on the inclination angle $i$ of galaxies, we also plot the mean values of H$\alpha$/H$\beta$ for galaxies with $i<30^{\circ}$ (nearly face-on) and $i>60^{\circ}$ (nearly edge-on). The inclination angle indeed has an impact on the value of H$\alpha$/H$\beta$, but the trend remains unchanged.

Since the gas in the central region of galaxies is dominated by \htwo~\citep{Leroy2008self}, the decrease of dust attenuation with decreasing SFR provides independent support of the trend for \htwo~shown in the left panel of Figure \ref{h2}. It is interesting to notice that the passive central disk galaxies with the lowest observable SFRs have H$\alpha$/H$\beta$ close to the intrinsic value of 3.1 (horizontal dash line) for a hard radiation field \citep{Osterbrock2006}. Since these quiescent galaxies with emission lines are dominated by LI(N)ERs \citep{Belfiore2017self,Guo2019self}, the H$\alpha$/H$\beta$ with an average value of $\sim$3.1 indicates that there is little dust and gas in the central region of these quiescent disk galaxies. 

As discussed in the Appendix, the main results presented in Figure \ref{hi} and Figure \ref{h2} can be greatly affected by inaccurate SFRs due to aperture corrections and/or extinction corrections. Therefore, we repeat our analysis by using the SFRs derived from the SED fitting of UV, optical and mid-IR bands \citep{Salim2018self}. As shown in Figure A1 and A2, all trends remain qualitatively the same as these in Figure \ref{hi} and Figure \ref{h2}.

%-----------------------------------------------------------------------------------------------------------------------------
\section{Summary and discussion}

In order to investigate the detailed quenching process with an internal physical origin, we studied the cold gas content in massive central disk galaxies in the local Universe. Our results show that massive quiescent central disk galaxies surprisingly have similar large amount of \hi~gas as star-forming ones. These galaxies are quenched because of their significantly reduced molecular gas and dust content and lower star formation efficiency.

The \hi~surface density in the outer regions of disk galaxies follows a homogenous radial profile when the radius is normalized by the diameter of the \hi~disk \citep{Broeils1997self,Wang2016self}. As mentioned above, quiescent central disk galaxies exhibit similarly symmetric characteristic double-horn \hi~profiles as star-forming systems, strongly suggesting that both galaxy types have regularly rotating \hi~disks. The almost constant \hi~gas mass of $\sim$$10^{10}\Msol$ (Figure \ref{h2}), across the entire observable range of SFR, corresponds to \hi~disks with radius of $\sim$30\,kpc \citep{Broeils1997self,Wang2016self}. Thus, the \hi~gas in quenched disks may be stored in an outer ring such as in the prototypical case of the S0 galaxy NGC 1543 \citep{Murugeshan2019self}.

These observational evidences suggest the following picture. Since the \hi~gas is distributed on very large scales \citep{Leroy2008self}, once the central \htwo~gas is consumed by star formation, expelled by outflows \citep{Fabian2012,Harrison2017a,Hopkins2014,Yuan2018self} or ionized/photo-dissociated by UV radiation from AGNs \citep{Fabian2012}, the timescale for the \hi~gas with high angular momentum in the outer disc to migrate inward may be very long in the absence of perturbations \citep{Renzini2018}. Therefore, during the quenching process, the rotationally supported outer \hi~disk remains largely unchanged. The SFR decreases driven by the decreasing \htwo~gas mass in the central region and progressively suppressed \htwo~star formation efficiency (cf. Figure \ref{h2}), with the gas remaining atomic rather than replenishing the star-forming molecular phase. The implications of these findings for the quenching of star formation in disk galaxies are further explored in \citet{Peng2019} and Zhang et al. (in preparation).

%---------------------------------------------------------------------------------------------------------------------
\acknowledgments

We thank Robert Kennicutt, Sandra Faber, Nick Scoville, Barbara Catinella, Renbin Yan and Jing Wang for useful discussions. We thank the ALFALFA team for providing the \hi~spectra of the ALFALFA extragalactic sample. Y.P. acknowledges National Key R\&D Program of China Grant 2016YFA0400702 and NSFC Grant No.\thinspace11773001. L.C.H. acknowledges National Key R\&D Program of China Grant 2016YFA0400702 and NSFC Grant No.\thinspace11473002 and 11721303. C.Z. acknowledges the NSFC Grant No.\thinspace11373009 and 11433008. R.M. acknowledges ERC Advanced Grant 695671 ``QUENCH'' and support by the Science and Technology Facilities Council (STFC). A.D. acknowledges support from the grants NSF AST-1405962, GIF I-1341-303.7/2016, and DIP STE1869/2-1 GE625/17-1. Q.G. acknowledges NSFC Grants No.\thinspace11573033, 11622325 and the Newton Advanced Fellowships.  F.M. acknowledges support from the INAF PRIN-SKA 2017 program 1.05.01.88.04. D.L. acknowledges the NSFC Grant No.\thinspace11690024 and 11725313. A.R. acknowledges support from an INAF/PRIN-SKA 2017 (ESKAPE-HI) grant.

%---------------------------------------------------------------------------------------------------------------------

\appendix

\section{Alternative SFR estimates}

The main results presented in Figure \ref{hi} and Figure \ref{h2} can be greatly affected by inaccurate SFRs, for instance due to inaccurate aperture corrections or extinction corrections. To address this concern, we repeat our analysis by using SFRs derived from the SED fitting of UV, optical and mid-IR bands \citep[GSWLC-M2 catalogue,][]{Salim2016a,Salim2018self}. The results are shown in Figure A1 and the left panel of Figure A2. The SFRs used in the right panel of Figure A2 are given by the xCOLD GASS catalogue \citep{Saintonge2017self} and are derived from the near-UV and mid-IR photometries by the technique described in \citet{Janowiecki2017self}. For both of these two alternative SFR estimates, the dynamical range of SFRs for quiescent disk galaxies becomes smaller, however, the general trends of these plots remain the same. For disk galaxies with SFRs well below the main-sequence ($\sim$1\,dex lower), the mean \hi~gas mass and \hi~detection ratio are still similar with those of star-forming galaxies as shown in Figure A1, while their \htwo~gas mass decreases by a factor of $\sim$10 during quenching as shown in Figure A2. The drop of star formation efficiency ($\varepsilon_{\rm H_2}$) becomes smaller compared to that of the MPA-JHU SFR.

\begin{figure*}[htbp]
    \begin{center}
       \includegraphics*[width=180mm]{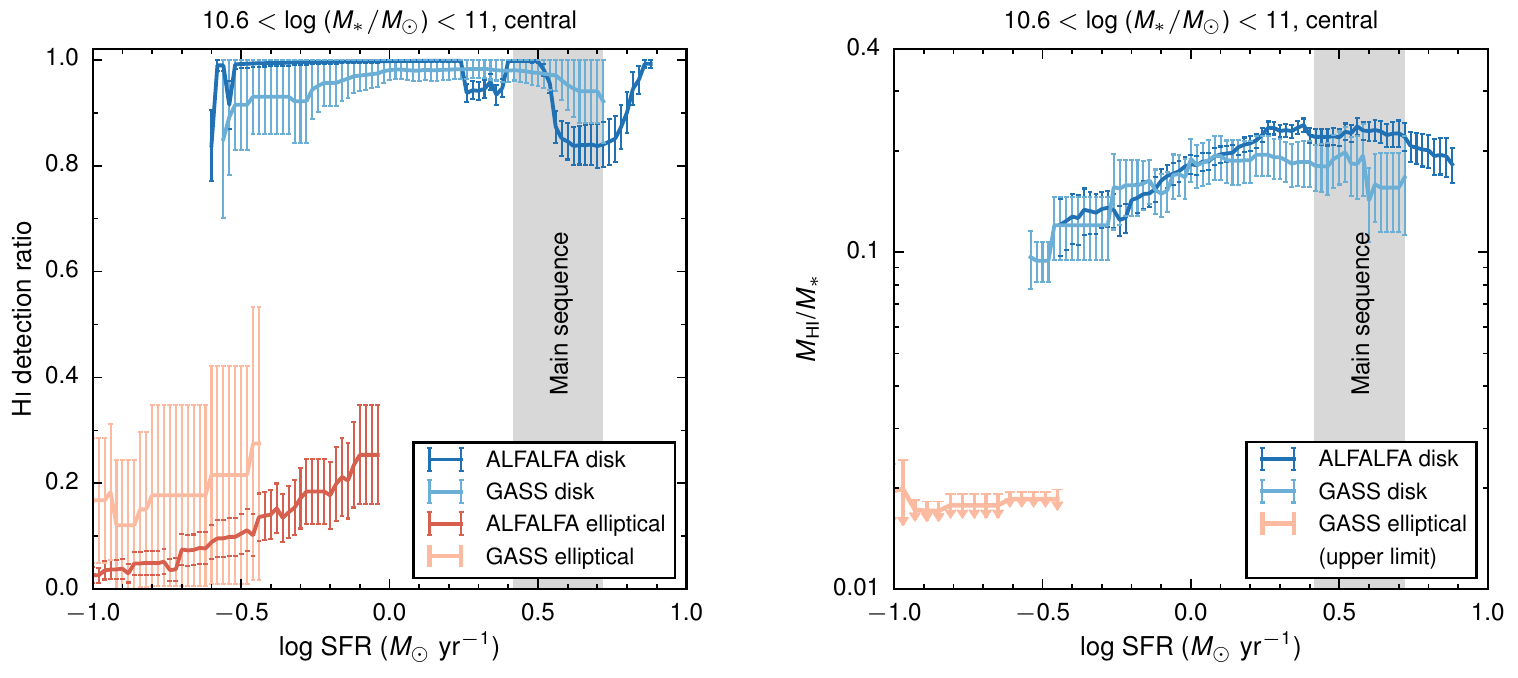}
    \end{center}
\noindent\textbf{Figure A1.} \hi~detection ratio and \hi~gas mass to stellar mass ratio (\mhi/$M_*$) for central disk galaxies (blue lines) and central elliptical galaxies (red lines) in the stellar mass range of $10^{10.6} - 10^{11}\Msol$, plotted by using the SFRs derived from the SED fitting of UV, optical and mid-IR bands \citep{Salim2018self}. All labels in this plot are the same with Figure \ref{hi}.
 \label{Fig_S18M}
\end{figure*}

\begin{figure*}[htbp]
    \begin{center}
       \includegraphics*[width=180mm]{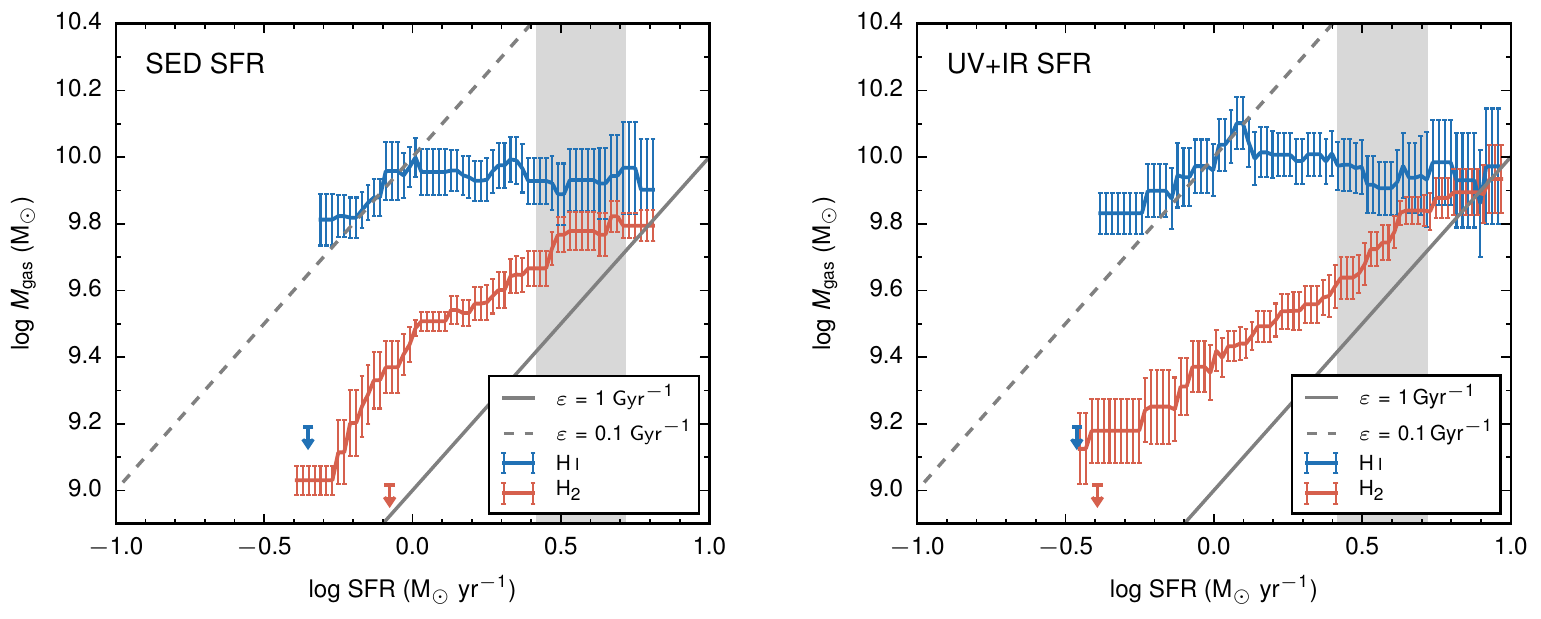}
    \end{center}
\noindent\textbf{Figure A2.} The average atomic gas mass (in blue) and molecular gas mass (in red) as a function of SFR for central disk galaxies in the stellar mass range of $10^{10.6} - 10^{11}\Msol$, plotted by using two different SFR estimates. The SFRs used in the left panel are derived from the SED fitting of UV, optical and mid-IR bands \citep{Salim2018self}. The SFRs used in the right panel are given by the xCOLD GASS catalogue \citep{Saintonge2017self} and they are derived from the near-UV and mid-IR photometries by the technique described in \citet{Janowiecki2017self}. All labels in this plot are the same with Figure \ref{h2}.
 \label{sfr}
\end{figure*}

\end{document}